\documentclass[journal]{IEEEtran}
\IEEEoverridecommandlockouts
\usepackage{cite}
\usepackage{amsmath,amssymb,amsfonts}
\usepackage{algorithmic}
\usepackage{graphicx}
\usepackage{textcomp}
\usepackage{xcolor}
\usepackage{color}
\usepackage{verbatim}
\usepackage{bm} 
\usepackage[T1]{fontenc}
\usepackage{array}
\usepackage{algorithm,algorithmic}
\usepackage{booktabs}
\newcommand{\mv}[1]{\mbox{\boldmath{$ #1 $}}}

\def\BibTeX{{\rm B\kern-.05em{\sc i\kern-.025em b}\kern-.08em
    T\kern-.1667em\lower.7ex\hbox{E}\kern-.125emX}}
\begin{document}

\title{Power-Measurement-Based Channel Estimation for Beyond Diagonal RIS}
\author{Yijie Liu, Weidong Mei, \IEEEmembership{Member, IEEE}, He Sun, \IEEEmembership{Member, IEEE}, Dong Wang, \IEEEmembership{Student Member, IEEE}, Zhi Chen, \IEEEmembership{Senior Member, IEEE}\vspace{-6pt}
\thanks{
Yijie Liu is with the School of Information and Communication Engineering, University of Electronic Science and Technology of China, Chengdu 611731, China. Weidong Mei, Dong Wang and Zhi Chen are with the National Key Laboratory of Wireless Communications, University of Electronic Science and Technology of China, Chengdu 611731, China. (e-mails: yijieliu@std.uestc.edu.cn; wmei@uestc.edu.cn; DongwangUESTC@outlook.com; chenzhi@uestc.edu.cn).

He Sun is with the Department of Electrical and Computer Engineering, National University of Singapore, Singapore 117583 (e-mail: sunele@nus.edu.sg). 
}}
\maketitle

\begin{abstract}
Beyond diagonal reconfigurable intelligent surface (BD-RIS), with its enhanced degrees of freedom compared to conventional RIS, has demonstrated notable potential for enhancing wireless communication performance. However, a key challenge in employing BD-RIS lies in accurately acquiring its channel state information (CSI) with both the base station (BS) and users. Existing BD-RIS channel estimation methods rely mainly on dedicated pilot signals, which increase system overhead and may be incompatible with current communication protocols. To overcome these limitations, this letter proposes a new single-layer neural network (NN)-enabled channel estimation method utilizing only the easily accessible received power measurements at user terminals. In particular, we show that the received signal power can be expressed in a form similar to a single-layer NN, where the weights represent the BD-RIS's CSI. This structure enables the recovery of CSI using the backward propagation, based on power measurements collected under varying training reflection coefficients. Numerical results show that our proposed method can achieve a small normalized mean square error (NMSE), particularly when the number of training reflections is large.\vspace{-3pt}
\end{abstract}\vspace{-3pt}
\vspace{-8pt}
\section{Introduction}
Reconfigurable intelligent surface (RIS) technology is considered a key enabler for 6G development and a critical component of future wireless communication systems. A RIS is typically composed of numerous passive reflecting elements, each capable of independently adjusting its reflection patterns for various purposes, such as signal enhancement, interference cancellation, and target detection \cite{renzo2020smart,zheng2022survey}. Recently, a more advanced RIS architecture, known as beyond-diagonal RIS (BD-RIS), has gained significant attention in the realm of wireless communications. Compared to conventional RIS employing a diagonal reflection matrix, BD-RIS allows for a non-diagonal reflection matrix, thereby providing more degrees of freedom for electromagnetic wave manipulation and channel reshaping \cite{li2023beyond}. Several studies have proposed a variety of optimization technologies to design such non-diagonal reflection matrices. Interested readers are referred to \cite{li2024reconf,li2024coordinated,li2022ris} and the references therein.

In RIS- and BD-RIS-aided communication systems, channel estimation is essential for performance optimization. Depending on the RIS configuration, existing methods fall into two categories \cite{zheng2022survey}. For semi-passive RIS, additional sensing components are integrated into RIS to enable separate estimation of the user–RIS and base station (BS)-RIS channels. For fully passive RIS, separate estimation is not possible, but the cascaded BS-RIS-user channel can be acquired via pilot signals.

\begin{figure}[t]
\centerline{\includegraphics[width=0.64\linewidth]{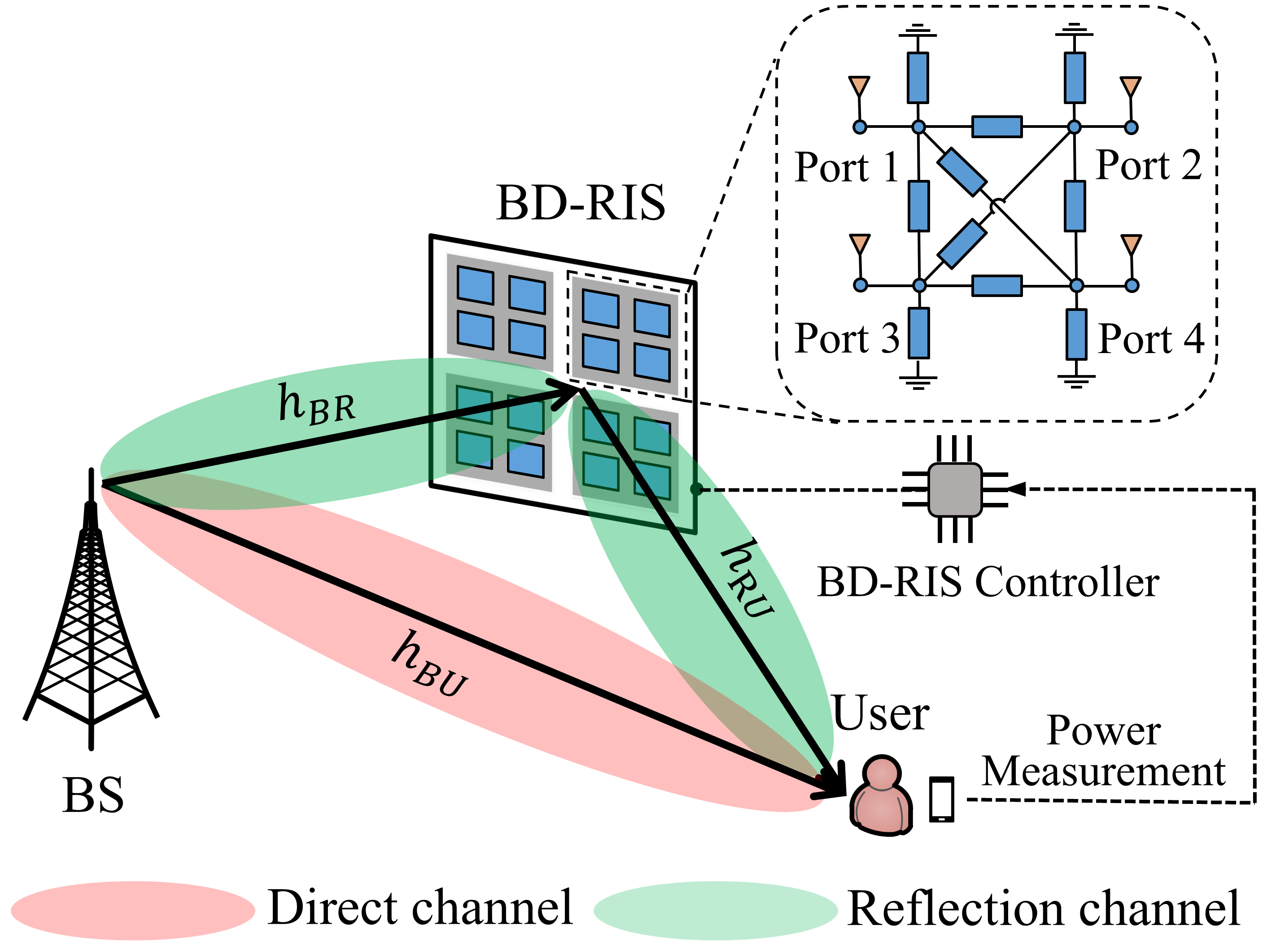}}
\caption{BD-RIS aided communication system.}
\label{SysMod}
\vspace{-8pt}
\end{figure}

Although the above methods are effective for conventional RIS in general, they may encounter difficulty if applied to BD-RIS due to the more complex reflection models. Particularly, with their non-diagonal reflection matrices, the cascaded channels for BD-RISs have a much higher dimension and may not be explicitly expressed compared to those for conventional RISs. A handful of recent studies have explored the channel estimation methods for BD-RISs \cite{li2024channel,almeida2024channel}. In \cite{li2024channel}, a baseline least squares (LS) receiver was developed for BD-RIS channel estimation, utilizing an orthogonal BD-RIS training matrix design subject to the physical constraints of BD-RIS. The authors in \cite{almeida2024channel} introduced a tensor decomposition method for BD-RIS channel estimation, where a set of BD-RIS training reflection patterns (TRPs) were devised. However, both of the above methods fail to account for the symmetric structural constraints for each block of the BD-RIS's block-diagonal reflection matrix, as imposed by its impedance network and other hardware constraints.
% symmetric structure of each block within the block-diagonal reflection matrix of the BD-RIS, which arises as an inherent constraint imposed by the topology of the underlying reconfigurable impedance network.} \textcolor{blue}{In addition, hardware imperfections introduce nonlinear distortions that degrade the accuracy of pilot-based channel estimation \cite{reviewer1_1,reviewer1_2}. More critically, reliance on dedicated pilot signals not only increases training overhead but also clashes with existing protocols, which are designed for direct user-BS channel estimation without BD-RIS support.}

To tackle the above limitations, we investigate a more practical channel estimation method for BD-RISs based on {\it power measurements at user terminals}, which are easily accessible in current cellular networks, as shown in Fig.\,\ref{SysMod}. Motivated by the recently proposed single-layer neural network (NN)-based method for conventional RIS in our prior work \cite{sun2023user,sun2024power}, we extend it to the more general and challenging scenario with a BD-RIS in this letter. Specifically, we show that the received signal power at each user can still be expressed in a form similar to a single-layer NN despite the non-diagonal structure of the BD-RIS, where the weights represent its related CSI. This structure thus enables the recovery of CSI using backward propagation, based on power measurements collected under varying TRPs. Furthermore, to reduce the training overhead, we propose a new TRP selection scheme by minimizing the maximum correlation among the selected TRPs. Numerical results demonstrate that our proposed method achieves a low normalized mean square error (NMSE), and the proposed TRP selection scheme outperforms random TRP selection. Moreover, the number of required TRPs increases with the number of groups of the BD-RIS.

\textit{Notations:} For any scalar/vector/matrix, $(\cdot)^*$, $(\cdot)^T$, and $(\cdot)^H$ respectively denote its conjugate, transpose and conjugate transpose. The Kronecker product of two matrices $\mv{A} \in \mathbb{R}^{m \times n}$ and $\mv{B} \in \mathbb{R}^{p \times q}$ is denoted by $\mv{A} \otimes \mv{B}$, which results in a block matrix of size $mp \times nq$.\vspace{-6pt}

\begingroup
\allowdisplaybreaks
\section{System Model and Problem Formulation}
As depicted in Fig.\,\ref{SysMod}, we consider a narrowband communication system from a single-antenna BS\footnote{Note that to ensure the plug-and-play deployment of the BD-RIS, the single-antenna BS can be equivalently viewed as a multi-antenna BS serving a user with an already optimized transmit beamforming.} to a single-antenna user\footnote{Our proposed channel estimation method can also be extended to multi-user scenarios by applying it independently to each user in parallel.}, with the aid of a BD-RIS equipped with $N$ reflecting elements. Unlike the conventional RIS with a diagonal reflection matrix,  BD-RIS can be modeled as $N$ reflecting elements connected to a reconfigurable impedance network of $N$ ports, with a scattering matrix denoted as $\mv{\Theta}\in\mathbb{C}^{N\times N}$. To maximize the reflected signal power of each BD-RIS and ease the hardware implementation, we assume that all ports of the reconfigurable impedance network are lossless. Let $\mv{X}\in\mathbb{R}^{N\times N}$  and $\mv{Z}\in\mathbb{C}^{N\times N}$ denote the reactance matrix and its associated imaginary impedance matrix, respectively, with $\mv{Z}=j\mv{X}$.  As such, the reconfigurable impedance network can be characterized as 
\begin{align}
\mv{\Theta}=(j\mv{X}+Z_0\mv{I})^{-1}(j\mv{X}-Z_0\mv{I}),\label{eq0}
\end{align}
where $Z_0$ represents the reference impedance, typically set to $50 \Omega$, $\mv{I}$ denotes the identity matrix of size $N$\cite{shen2022model}.

For a group-connected BD-RIS, the reflection matrix is block-diagonal and the $N$ elements can be divided into $K$ groups, each of which includes $N_0=N/K$ reflecting elements. $N_0$ is also known as the group size of BD-RIS. Notably, a BD-RIS reduces to a conventional RIS if $K=N$ or $N_0=1$. Let \( k \in \mathcal{K}\ \triangleq \{1, 2, \dots, K\}\) denote the set of groups. Thus, for the $k$-th group, it can be modeled as an $N_0$-port fully connected reconfigurable impedance network with an impedance matrix of $Z_0$. 
Let the reflection matrix of the $k$-th group be denoted by $\mv{\Theta}_k\in \mathbb{C}^{N_0\times N_0}$. Then, the reflection matrix of the whole BD-RIS is given by
\begin{equation}
\mv{\Theta}=\text{diag}(\mv{\Theta}_1, \dots, \mv{\Theta}_k,\dots, \mv{\Theta}_K), k\in\mathcal{K}.\label{eq1}
\end{equation}
Note that the physical structure of a BD-RIS leads to the conjugate symmetry property of $\mv{\Theta}_k$ \cite{shen2022model}. Accordingly,  we have
\vspace{-4pt}
\begin{equation}
\mv{\Theta}_k=(j\mv{X}_{k}+Z_0\mv{I})^{-1}(j\mv{X}_{k}-Z_0\mv{I}),k\in\mathcal{K},\label{eq2}
\end{equation}
where $\mv{X}_k\in \mathbb{C}^{N_0\times N_0}$ 
denotes the reactance matrix of the reconfigurable
impedance network in the $k$-th group.  

In this paper, we consider quasi-static block-fading channels and focus on a specific fading block e.g., the length of a slot specified in Long Time Evolution (LTE) \cite{LTE}, within which all channel coefficients remain constant. Let the channel from the BS to the user, that from the BS to the BD-RIS, and that from the BD-RIS to the user be denoted as $\mv{h}_{BU}\in \mathbb{C}$, $\mv{h}_{BR}\in \mathbb{C}^{N\times1}$, $\mv{h}_{RU}\in \mathbb{C}^{N\times1}$, respectively. Hence, the overall channel from the BS to the user (subsuming the transmit power) is given by
\begin{equation}\label{eq3}
g=\sqrt{P}({h}_{BU}+\mv{h}_{RU}^H\mv{\Theta} \mv{h}_{BR}),
\end{equation}
where \(P\) is the transmit power of the BS. Let $s$ denote the transmitted symbol (pilot or data) at the BS with $\vert s \vert^2=1$. Hence, the noiseless received signal at the user is given by
\begin{equation}
y=gs.
\end{equation}
Accordingly, the received signal power is
\begin{equation}\label{rsrp}
\eta = \lvert g \rvert^2.
\end{equation}

It is observed that the received signal power in \eqref{rsrp} can be maximized by adjusting the BD-RIS reflection matrix if the perfect CSI, i.e., $\mv{h}_{RU}$ and $\mv{h}_{BR}$, is available. However, perfect CSI is generally difficult to obtain in practice, especially for BD-RIS. Moreover, the existing channel estimation technique for BD-RIS shows incompatibility with the existing cellular protocol. As such, we propose an efficient single-layer NN-enabled method to estimate the CSI required with user power measurements only. \vspace{-6pt}

\section{BD-RIS Channel Estimation Using a Single-Layer Neural Network}
Before presenting the proposed method for BD-RIS, we first review the existing single-layer NN-based method designed for conventional RIS (or single-group BD-RIS). For an RIS consisting of $N$ reflecting elements, we denote its reflection matrix as $\mv{\Theta} = \text{diag}(e^{j\theta_1},...,e^{j\theta_N})$. Then, the end-to-end BS-user channel in \eqref{eq3} can be simplified as
\begin{equation}
g=\sqrt{P}(h_{BU}+\mv{\bar{\lambda}}^H\mv{\bar{\xi}})=\mv{\lambda}^H\mv{\xi},\label{add}
\end{equation}
where $\mv{\bar{\lambda}}^H=[e^{j\theta_1},...,e^{j\theta_N}]$, $\mv{\lambda}^H=[1,\mv{\bar{\lambda}}^H]$, $\mv{\bar{\xi}}=\text{diag}(\mv{h}_{RU}^H)\mv{h}_{BR}$ and 
$\mv{\xi}^H=\sqrt{P}[h^*_{BU},\mv{\bar{\xi}}^H]$.
Nevertheless, the reflection matrix of a BD-RIS is no longer a diagonal matrix but a block diagonal matrix. The effective channel in \eqref{eq3} cannot be written as (\ref{add}). Hence, the existing single-layer NN-based method for RIS cannot be directly applied to BD-RIS. To tackle this challenge, we recast the BD-RIS channel in (\ref{eq3}) into a more tractable form, as presented next.\vspace{-8pt}

\subsection{Reformulation of BD-RIS Channel Model}
For \eqref{eq3}, by utilizing the vec-Kronecker product property, i.e., \(\text{vec}(\mv{ABC}) = (\mv{C}^T \otimes \mv{A}) \text{vec}(\mv{B})\) for any arbitrary matrices \(\mv{A}\), \(\mv{B}\), and \(\mv{C}\), we have
\begin{equation}
\text{vec}(\mv{h}_{RU}^H \mv{\Theta}\mv{h}_{BR}) = (\mv{h}_{BR}^T \otimes \mv{h}_{RU}^H) \text{vec}(\mv{\Theta}).\label{eq5}
\end{equation}
The conjugate transpose property of the Kronecker product, \((\mv{A} \otimes\mv{B})^H = \mv{A}^H \otimes \mv{B}^H\), further implies that
\begin{equation}
 (\mv{h}_{BR}^T \otimes \mv{h}_{RU}^H) \text{vec}(\mv{\Theta})=(\mv{h}_{BR}^* \otimes \mv{h}_{RU})^H \text{vec}(\mv{\Theta}).\label{eq6}
\end{equation}
From the Kronecker product property, \(\text{vec}(\mv{ab}^T) = \mv{b} \otimes \mv{a}\), we have
\begin{equation}
(\mv{h}_{BR}^* \otimes \mv{h}_{RU})^H \text{vec}(\mv{\Theta}) = \text{vec}(\mv{M})^H \text{vec}(\mv{\Theta}), \label{eq7}
\end{equation}
where $\mv{M} = \mv{h}_{RU} \mv{h}_{BR}^H$. Combining equations (\ref{eq5}), (\ref{eq6}) and (\ref{eq7}), we can show that
\begin{equation}
\mv{h}_{RU}^H \mv{\Theta} \mv{h}_{BR} = \text{vec}(\mv{\Theta}^*)^H\text{vec}(\mv{M}^*)=\mv{\hat{v}}^H \mv{\hat{h}},\label{eq4}
\end{equation}
 where \(\mv{\hat{v}} = \text{vec}(\mv{\Theta}^*)\) and \(\mv{\hat{h}} = \text{vec}(\mv{M}^*)\). 
Thus, the channel in \eqref{eq3} is equivalently recast as
\begin{equation}
g = \sqrt{P} ( {h}_{BU} + \mv{\hat{v}}^H \mv{\hat{h}}).\label{eq8}
\end{equation}
%\subsection{Problem Formulation}\label{AA}

Let $\mv{h}^T=\sqrt{P}[h_{BU}, \hat{\mv{h}}^T]$, and $\mv{v}^H=[1,\hat{\mv{v}}^H]$. The channel in (\ref{eq8}) can be simplified as
\begin{equation}
g = \mv{v}^H\mv{h},\label{eq9}
\end{equation}
where $\mv{v}\in\mathbb{C}^{(N^2+1)\times1}$ and $ \mv{h}\in\mathbb{C}^{(N^2+1)\times1}$. Note that we can view $\mv{h}$ as the cascaded channel for BD-RIS. Compared to the conventional RIS cascaded channel, its dimension increases from $N$ to $N^2+1$. For any given BD-RIS reflection $\mv{v}$, the received signal power is given by
\begin{equation}
\eta(\mv{v}) = \lvert \mv{v}^H\mv{h} \rvert^2.\label{eq11}
\end{equation}

As such, the received signal power is expressed in a form similar to a single-layer NN. Noting that some elements of the BD-RIS reflection vector \( \mv{v} \) are zero due to the block diagonal structure, we can further simplify (\ref{eq11}) by eliminating zero elements from both \( \mv{v} \) and \( \mv{h} \). Based on (\ref{eq1}), we can rewrite vector \( \mv{v} \) as
\begin{equation}
\mv{\bar{v}}^H=[1,\text{vec}(\mv{\Theta}_1)^T,\dots,\text{vec}(\mv{\Theta}_k)^T,\dots,\text{vec}(\mv{\Theta}_K)^T], \; k\in \mathcal{K}. \label{eq12}
\end{equation}

As \( \mv{\hat{h}} \) and \( \mv{\hat{v}} \) have the same dimension, the zero elements of \( \mv{\hat{v}} \) will nullify the corresponding entries in \( \mv{\hat{h}} \) in their inner product. Following the similar process of constructing \( \mv{\bar{v}} \), we can construct a block diagonal version of $M$ as
\begin{equation}
\hat{\mv{M}}=\text{diag}(\mv{M}_1, \dots, \mv{M}_k, \dots, \mv{M}_K), \quad k\in \mathcal{K},\label{eq13}
\end{equation}
with $\mv{M}_k \in \mathbb{C}^{N_k \times N_k}$. By defining
\[
\mv{\bar{h}}^T=[\mv{h}_{BU},\text{vec}^T(\mv{M}_1), \dots, \text{vec}^T(\mv{M}_k), \dots, \text{vec}^T(\mv{M}_K)],
\]
we can recast \eqref{eq8} as
\begin{equation}
g = \mv{\bar{v}}^H\mv{\bar{h}},\label{eq15}
\end{equation}
where $\mv{\bar{v}}\in\mathbb{C}^{(N_{0}N+1)\times1}, \mv{\bar{h}}\in\mathbb{C}^{(N_{0}N+1)\times1}$. Compared to (12), it is noted that the dimensions of $\mv{v}$ and $\mv{h}$, i.e., $N^2+1$, reduce to those of $\bar{\mv{v}}$ and $\bar{\mv{h}}$ i.e., $N_{0}N+1$. Therefore, the received signal power in (\ref{eq11}) can be rewritten as
\begin{equation}
\eta(\mv{\bar{v}}) = \lvert \mv{\bar{v}}^H\mv{\bar{h}} \rvert^2 = \mv{\bar{v}}^H\mv{\bar{h}}\mv{\bar{h}}^H\mv{\bar{v}}=\mv{\bar{v}}^H\mv{G}\mv{\bar{v}}\label{eq16},
\end{equation}
where $\mv{G}=\mv{\bar{h}}\mv{\bar{h}}^H$ denotes the cascaded channel autocorrelation matrix for BD-RIS.

Based on \eqref{eq16}, we can adopt a single-layer NN-based channel estimation approach to estimate $\mv{\bar{h}}$ or $\mv{G}$. Specifically, let $\mv{\bar{v}}$ and $\eta(\mv{\bar{v}})$ represent the input and output data, respectively. As such, we can regard $\mv{\bar{h}}$ as the weight, with the activation function in the output layer being the squared amplitude of $\mv{\bar{v}}^H\mv{\bar{h}}$.

\newtheorem{remark}{Remark}
\begin{remark}
The received signal expression in (\ref{eq15}) is reminiscent of the classical LS-based channel estimation method, where a set of TRPs, i.e., $\bar{\mv{v}}$, is generated to estimate the cascaded channel $\bar{\mv{h}}$. However, this method may fail in the context of BD-RIS due to the structural constraint in (\ref{eq0}), which prevents $\bar{\mv{v}} $ spanning the entire $(N_0N+1)$-dimensional space. Moreover, it is inconsistent with the power-measurement-based channel estimation framework considered in this letter.
\end{remark}
\vspace{-6pt}

\subsection{TRP Selection}
Before training the neural network, we need to properly generate a set of TRPs as the input. Let $\mathcal{D}$ denote the set of inputs, with $\lvert {\cal D} \rvert=D$. Denote by $\bar{\mv{v}}_d$ the $d$-th TRP. Then, we have
$\mathcal{D}=\{\mv{\bar{v}}_1, \mv{\bar{v}}_2, \dots, \mv{\bar{v}}_{n}\}$. To determine $\cal D$, we aim to minimize the maximum correlation between any two TRPs in $\cal D$, thereby reducing redundancy and improving training efficiency. The associated TRP selection problem can be formulated as
\begin{equation}\label{eq:TRP_opt}
    \min_{|{\mathcal{D}}|=D} 
    \max_{\substack{\bar{\mv{v}}_i,\bar{\mv{v}}_j\in\mathcal{D},\, i\neq j}}
    \big|\mathrm{Corr}(\bar{\mv{v}}_i, \bar{\mv{v}}_j)\big|,
\end{equation}
where $\mathrm{Corr}({\bar{\mv{v}}_i}, \bar{\mv{v}}_j)=\frac{\bar{\mv{v}}^H_i\bar{\mv{v}}_j}{\lvert \bar{\mv{v}}_i\rvert \cdot \lvert \bar{\mv{v}}_j \rvert}$ denotes the correlation coefficient between two TRPs $\bar{\mv{v}}_i$ and $\bar{\mv{v}}_j$.  
Since \eqref{eq:TRP_opt} is a combinatorial optimization problem and becomes intractable to tackle for a large value of $D$, we propose a sequential greedy selection algorithm to obtain a suboptimal solution to \eqref{eq:TRP_opt}. Specifically, we first generate a candidate TRP pool, denoted as $\cal C$, with $\lvert {\cal C} \rvert=C \gg D$. Then, we sequentially select TRPs from $\cal C$, such that the newly selected TRP at each step has the minimum correlation with all previously selected TRPs. The main procedures are summarized in Algorithm~\ref{alg:greedy_selection_condensed}. 

\begin{algorithm}[!ht]
    \renewcommand{\algorithmicrequire}{\textbf{Input:}}
    \renewcommand{\algorithmicensure}{\textbf{Output:}}
    \caption{Greedy TRP Selection Algorithm}
    \label{alg:greedy_selection_condensed}
    \begin{algorithmic}[1]
        \REQUIRE  Candidate TRP pool ${\cal C}$ and desired TRP number $D$
        \ENSURE  TRP set $\mathcal{D}$
        \STATE Select any $\bar{\mv v}_1\in\mathcal{C}$; 
        \STATE Update $\mathcal{D}\gets\{\bar{\mv v}_1\}$ and $\mathcal{C}\gets\mathcal{C}\setminus\{\bar{\mv{v}}_1\}$
        \FOR{$k=2$ \TO $D$}
            \STATE $\bar{\mv v}^* \!\gets\!\arg\min_{\bar{\mv v}_c\in\mathcal{C}}\max_{\,\bar{\mv v}_d\in\mathcal{D}}\bigl|\mathrm{Corr}(\bar{\mv v}_c,\bar{\mv v}_d)\bigr|$
            \STATE Update $\mathcal{D}\!\gets\!\mathcal{D}\cup\{\bar{\mv v}^*\}$ and $\mathcal{C}\!\gets\!\mathcal{C}\setminus\{\bar{\mv v}^*\}$
        \ENDFOR
    \end{algorithmic}
\end{algorithm}\vspace{-18pt}

\subsection{Single-Layer NN-Enabled Channel Estimation}
With the selected TRP set $\mathcal{D}$ obtained from Algorithm 1, we now train the single-layer network. Based on (\ref{eq16}), the received signal power can be expressed as
\begin{equation}\label{actual}
\tilde{\eta}(\mv{\bar{v}}_d)=  \lvert \mv{\bar{v}}_d^H\mv{\bar{h}} \rvert^2 + e_d,
\end{equation}
where $\bar{\mv{v}}_d$ denotes the $d$-th TRP and $e_d$ denotes the measurement error in $\mathcal{D}$.\footnote{Note that potential nonlinear distortion in the BS’s radio-frequency (RF) front-end during data/pilot transmission may affect the accuracy of the power measurements. For simplicity, we assume that such nonlinear effects have been well mitigated \cite{nonlinear}, with any residual distortion absorbed into $e_d$.} Hence, we can train the single-layer NN to minimize the mean squared error (MSE) between its output \eqref{eq16} and the actual power measurements in \eqref{actual} and estimate the cascaded channel $\bar{\mv{h}}$ as the NN weights accordingly. {Note that the power measurements in \eqref{actual} are referred to as reference signal received power (RSRP) in LTE \cite{LTE}, which can be measured and fed back to the BS via dedicated control links. In addition, to ensure correct matching between the power measurements and their corresponding TRPs, the BD-RIS controller should be time-synchronized with the user, which can be achieved using the primary and secondary synchronization signals in LTE \cite{LTE}. }

\begin{figure}[t]
\centerline{\includegraphics[width=1.0\linewidth]{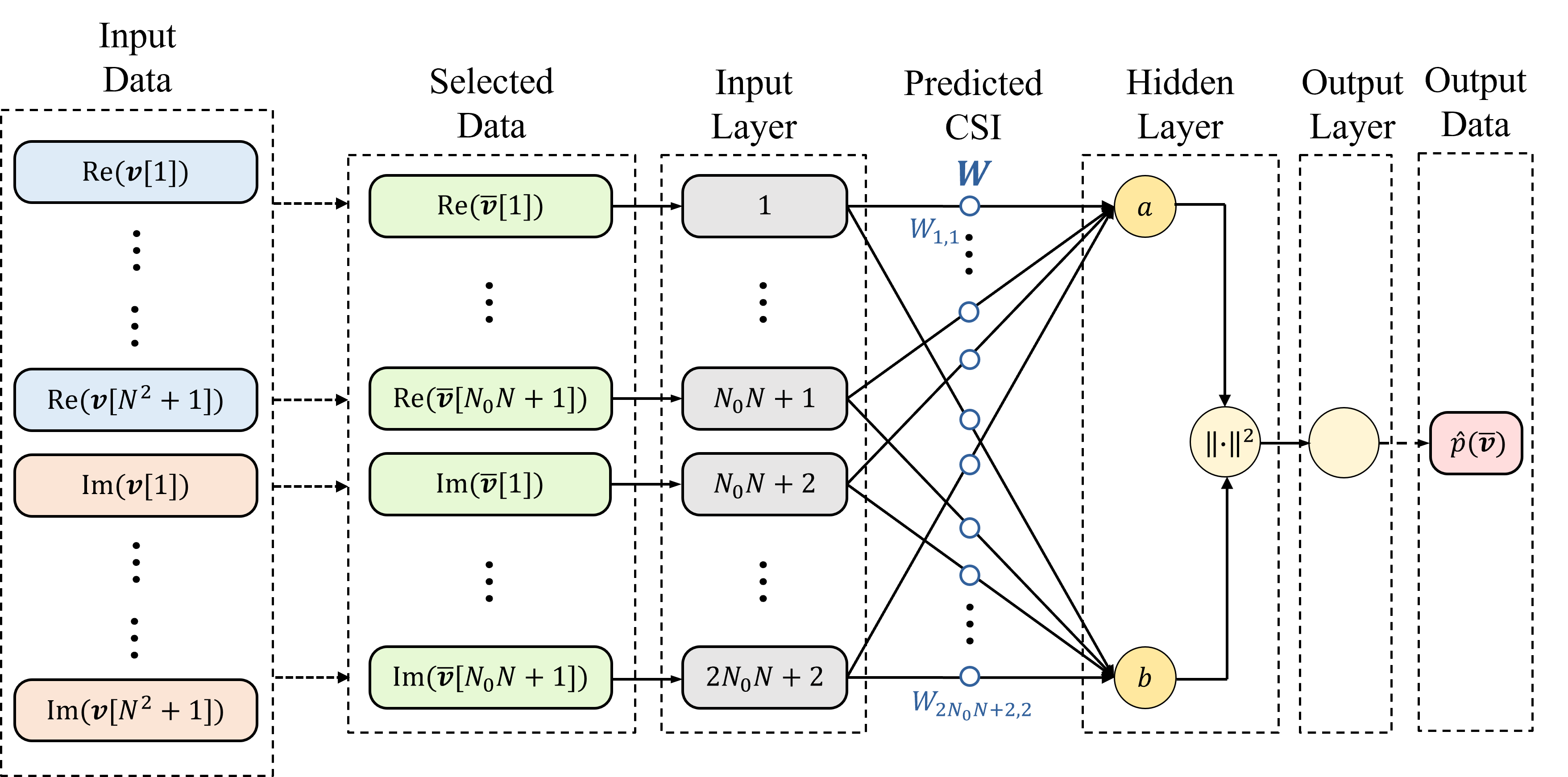}}
\caption{Single-layer NN architecture for the user.}
\label{fig2}
\vspace{-8pt}
\end{figure}
Although this training is effective, both the input and weight are complex numbers that are not convenient for weight update in a single-layer NN. To tackle this challenge, we need to convert the complex NN into a real-valued NN. To this end, we rewrite (\ref{eq16}) as
\begin{equation}
\eta(\mv{\bar{v}}) = \lvert \mv{\bar{v}}^H\mv{\bar{h}}\rvert^2=\Vert \mv{x}^T\mv{R} \Vert^2,\label{eq20}
\end{equation}
where $\mv{x}$ consists of the real and imaginary parts of $\mv{\bar{v}}$, i.e., $\mv{x}^T=[ \text{Re}(\mv{\bar{v}}^T), \text{Im}(\mv{\bar{v}}^T) ]$, and $\mv{R}$ denotes the real-valued cascaded channel, i.e.,  
\begin{equation}
\mv{R} = \left[\begin{array}{cc}
\text{Re}(\bar{\mv{h}}) & \text{Im}(\bar{\mv{h}}) \\
\text{Im}(\bar{\mv{h}}) & -\text{Re}(\bar{\mv{h}})
\end{array}\right]\in \mathbb{R}^{(2N_0N+2)\times2}.\label{eq21}
\end{equation}
Based on (\ref{eq20}), we can construct an equivalent single-layer NN in the real-number domain. As illustrated in Fig. 2, the input to this single-layer neural network is $\mv{x}$. Let $W_{i,j}$ represent the weight connecting the $i$-th input to the $j$-th neuron in the hidden layer, $i = 1, 2, ..., 2N_0N+2$ and $j = 1, 2$. The values of the two neurons in the hidden layer are given by
\begin{equation}
    [a\;\; b]
    =
    \mv{x}^T \mv{W},\label{a,b}
\end{equation}
where $\mv{W} \in \mathbb{R}^{(2N_0N+2) \times 2}$ denotes the weight matrix of this NN. Finally, the output of this NN is
\begin{equation}
    {\eta}_0(\mv{x}) = a^2 + b^2 = \left\lVert \mv{x}^T \mv{W} \right\rVert^2.
\end{equation}

To obtain the network weight matrix \( \mv{W} \), we randomly split both the input dataset \( \{\mv{x}_d\} \) (where \( \mv{x}_d^T = [\text{Re}(\mv{\bar{v}}_d^T), \text{Im}(\mv{\bar{v}}_d^T)] \)) and output dataset \( \{{\eta}_0(\mv{x}_d)\} \) into training and validation subsets. The first \( D_0 \) samples (\( D_0 < D \)) are used for training, with the remaining \( D-D_0 \) reserved for validation. The loss function for weight update is defined as   
\begin{equation}  
\mathcal{L}_{\boldsymbol{W}} = \frac{1}{D_0} \sum_{d=1}^{D_0}( \tilde{\eta}(\mv{\bar{v}}_d) - \eta_0(\mv{x}_d) )^2. 
\label{eq24}  
\end{equation}

Based on this loss function, we can use backward propagation to iteratively update the NN weights. Due to the space limit, the details of the gradient updates are omitted, while they can be found in our previous work \cite{sun2024power}. The NN training process terminates after $R$ iterations, and the weight matrix of the NN is determined as
\begin{equation}
\mathbf{\mv{W}}^* = \arg \min_{1 \leq t \leq R} \left( \sum_{d=D_0+1}^{D} \left({\eta}_{t}(\mv{x}_d) - \tilde\eta(\mv{\bar{v}}_d) \right)^2 \right),\label{eq30}
\end{equation}
based on the validation set, where ${\eta}_t(\mv{x}_d) = \left\| \mv{x}_d^T \mv{W}_{t} \right\|^2$ denotes the output of the NN after the $t$-th iteration, with $\mv{W}_t$ denoting the updated version of $\mv{W}$ after the $t$-th iteration. By this means, the complex-valued cascaded channel $\mv{\bar{h}}$ (subject to an unknown common phase \cite{sun2024power}) can be retrieved via $\mv{W}$. Note that the unknown common phase does not impact the recovery of the channel autocorrelation matrix $\mv{G}$ and the subsequent performance optimization for BD-RIS. However, it relaxes the constraints on the TRPs required to span the entire $(N_0N+1)$-dimensional space.

It can be shown that the complexity order of the single-layer NN architecture is \(\mathcal{O}(N_0N)\). It is also worth noting that, unlike conventional NN-based data prediction with unknown input-output relationships, our proposed single-layer NN adopts an expression that exactly matches the received signal power shown in (\ref{eq16}). This ensures its strong generalization capability for unseen data, as also demonstrated numerically in \cite{sun2024power}. \vspace{-8pt}

\renewcommand{\baselinestretch}{0.95}
\section{Numerical Results}
In this section, we provide numerical results to validate the efficacy of the proposed single-layer NN-based method in estimating the channel autocorrelation matrix for BD-RIS.\vspace{-12pt}

\subsection{Simulation Parameters}   
We consider a communication system with a BS, a BD-RIS, and a user as shown in Fig.\,\ref{SysMod}. The BS is assumed to be deployed at $(50, -200, 20)$ in meters (m), while the user's location is randomly generated within a square area with the coordinates of its four corner points given by $(0, 0, 0)$, $(10, 0, 0)$, $(10, 10, 0)$ and $(0, 10, 0)$, respectively. The BD-RIS is equipped with a uniform planar array (UPA) with $N = N_y \times N_z$ reflecting elements, with $N_y = N_z = 4$ and half-wavelength spacing between adjacent reflecting elements. The location of the reference point for the BD-RIS is set to $(-2, -1, 0)$. The BS-user channel follows Rayleigh fading, with the path loss given by $\beta_0 = 33 + 37\log_{10}(d_0)$ in dB, where $d_0$ is the distance from the BS to the user. The path losses for the BS-BD-RIS and BD-RIS-user channels are set as $\beta_1 = 30 + 20\log_{10}(d_1)$ and $\beta_2 = 30 + 20\log_{10}(d_2)$ in dB, respectively, where $d_1$ and $d_2$ are the BS-BD-RIS and BD-RIS-user distances, respectively. The BS transmit power is $P = 30$ dBm. \vspace{-9pt}

\subsection{Cosine Annealing-Based Learning Rate Scheduling} 
The single-layer NN may encounter convergence issues due to the sensitivity of the single-layer NN to parameter selection. To enhance convergence and stability, we introduce a cosine annealing-based learning rate scheduling strategy, where the learning rate $\mu$ dynamically decreases as training progresses.

Initially, a larger learning rate allows for faster convergence, while in the later stages, the learning rate decreases for finer adjustments. The cosine annealing schedule is defined as
\begin{equation}
\mu_t = \mu_{\min} + \frac{1}{2} \left( \mu_{\max} - \mu_{\min} \right) \left( 1 + \cos\left( \frac{T_t}{T_{max}} \pi \right) \right),\label{eq34}
\end{equation}
where $\mu_t$ represents the learning rate at iteration $t$, $\mu_{\min}$ and $\mu_{\max}$ are the minimum and maximum values of the learning rates, respectively, $T_t$ denotes the current iteration step, and $T_{\max}$ represents the total number of steps per cosine cycle.

With this strategy, the learning rate $\mu$ decays according to the cosine function and oscillates within a cycle, thereby improving both the exploration ability and the convergence of the training process. Moreover, the NN can quickly reduce the loss function in the early stages of training, while gradually stabilizing in the later stages, thus effectively mitigating the instability caused by the sensitivity to the parameters. \vspace{-9pt}

\subsection{Simulation Results} 
 To evaluate the performance of the single-layer NN-based BD-RIS channel estimation method, we compute the average NMSE between the estimated channel autocorrelation matrix and the actual one ($\mv{G}$). Denote by $L$ the total number of Monte Carlo simulations (for random realizations of channels and user locations) and let $\mathcal{L}=\{1,2,\cdots,L\}$, with $L=100$. Let $\mv{\hat{G}}_l$ denote the estimated channel autocorrelation matrix in the $l$-th simulation. Then, the NMSE achieved by the proposed method is calculated as
$\text{NMSE} = \frac{1}{L} \sum_{l \in \mathcal{L}} \frac{\|\hat{\mv{G}}_{l} - \mv{G}\|_F^2}{\|\bm{G}\|_F^2}$.

\begin{figure}[t]
\centerline{\includegraphics[width=0.75\linewidth]{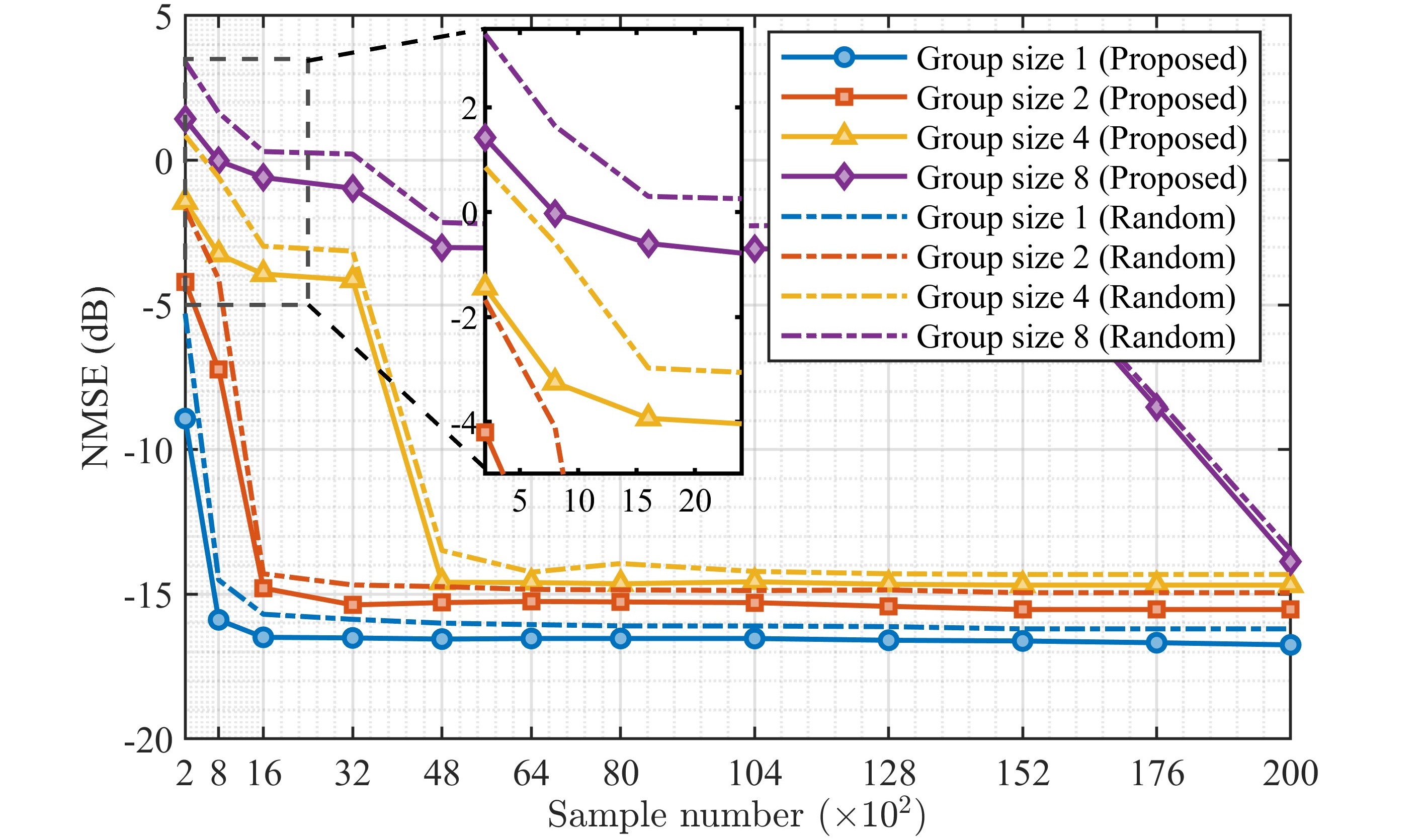}}
\caption{NMSE versus the TRP number, $D$.}
\label{fig3}
\vspace{-6pt}
\end{figure}
\begin{figure}[t]
\centerline{\includegraphics[width=0.75\linewidth]{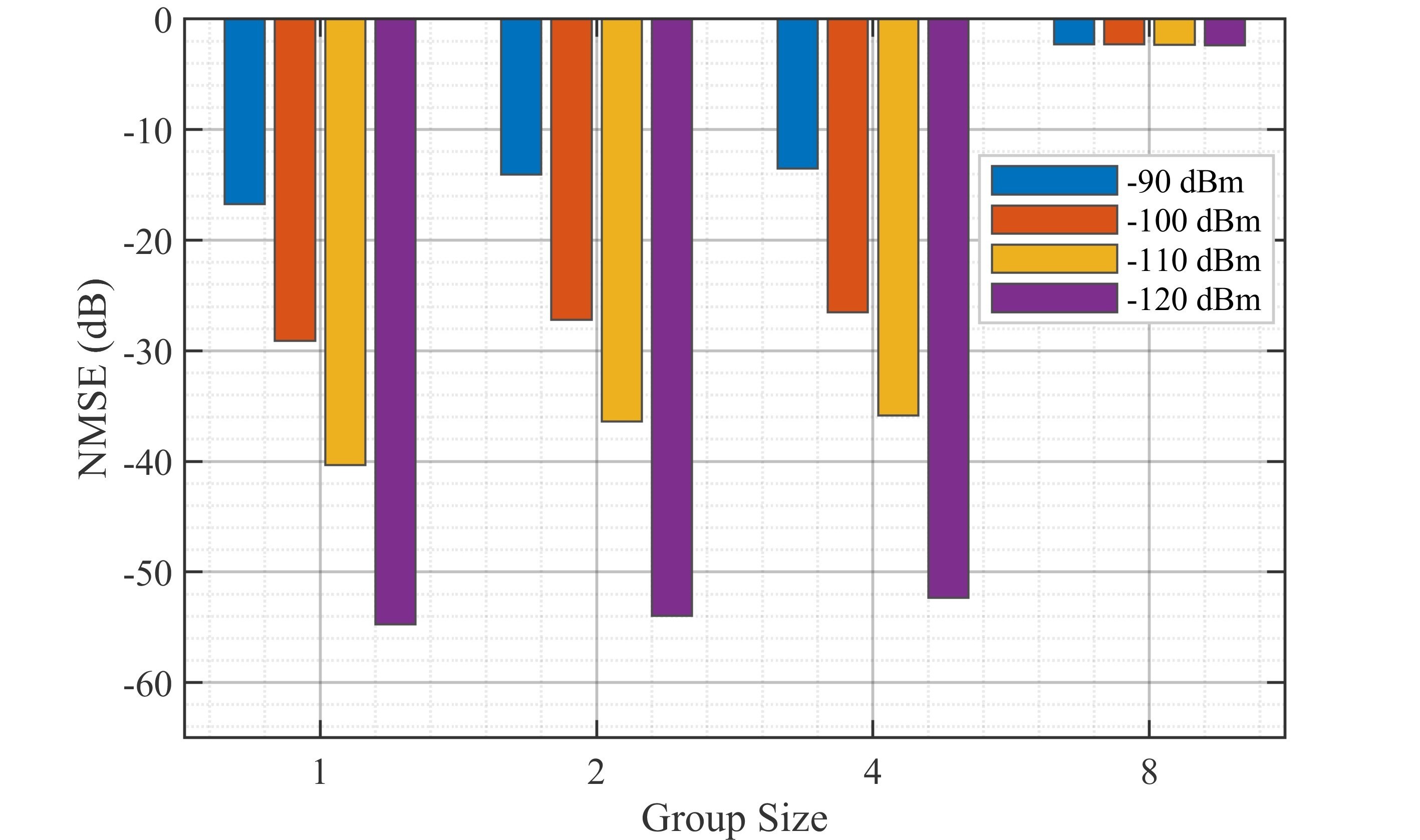}}
\caption{NMSE versus noise power under different group sizes.}
\label{fig4}
\vspace{-9pt}
\end{figure}
Fig.\,\ref{fig3} shows the NMSE by the proposed BD-RIS channel estimation method versus the number of TRPs (i.e., $D$) and group sizes (i.e., $N_0$) under two different TRP selection schemes, i.e., the proposed scheme and random TRP selection. The noise power in the power measurement is set to $-90$ dBm. It is observed that as the sample size increases, the NMSEs for all considered group sizes decrease. Nonetheless, the decreasing rate varies for different group sizes. The larger the group sizes, the higher NMSE for a given number of samples. This is expected, as the total number of
channel parameters to be estimated is given by $2N_0N +2$, which linearly increases with $N_0$ for a given BD-RIS element number $N$. It is also observed that our proposed TRP selection scheme yields a lower NMSE compared to the random selection, especially for a small sample number or a large group size.

Next, we fix the sample number as 8000 and show the NMSE by our proposed scheme by varying the noise power in the power measurement from $-120$ dBm to $-90$ dBm in Fig.\,\ref{fig4}. It is observed from Fig.\,\ref{fig4} that as the group size is 1, 2, and 4, a higher noise power results in a higher NMSE, as expected. While for a given noise power, it is observed that increasing the group size only slightly increases the NMSE despite a larger number of channel parameters to be estimated. This observation implies that the proposed single-layer NN-based channel estimation method is effective for BD-RIS even with a large group size, provided that the number of training samples is sufficient. For example, with only 8000 training samples, the NMSE for a group size of 8 cannot achieve its minimum NMSE as depicted in Fig.\,\ref{fig3}.\vspace{-6pt}

% \textcolor{blue}{Furthermore, we have provided simulation results of the normalized generalization error (NGE) obtained on an unseen test set with $U$ samples to investigate the generalization of our networks. Specifically, the NGE is defined as the expected value of the
% loss function, i.e., 
% \begin{equation}
%     \text{NGE}_t = \frac{1}{U} \sum_{u=1}^{U}\frac{( \tilde{\eta}(\mv{\bar{v}}_u) - \eta_0(\mv{x}_u) )^2}{ \tilde{\eta}(\mv{\bar{v}}_u)^2},
% \end{equation}
% where $\text{NGE}_t$ denotes the NGE after the $t$-th iteration. Fig. \ref{fig5} shows the relationship between NGE and training iteration rounds while $D_0=U=1600$, i.e., with 1600 samples for training and 800 unseen samples for testing. The noise power is set to $-90$ dBm. It is observed that the NGE decreases rapidly with increasing training rounds and cosine annealing-based learning rate scheduling strategy can be well applied to different group size scenarios. These results verified that our proposed algorithm has a good generalization capability.}

% \begin{figure}[t]
% \centerline{\includegraphics[width=0.9\linewidth]{Conference-LaTeX-template_10-17-19/generalization2.jpg}}
% \caption{NGE versus NN training iteration rounds}
% \label{fig5}
% \vspace{-6pt}
% \end{figure}
\section{Conclusion}
 In this letter, we proposed a single-layer NN-based channel estimation method for BD-RISs, by recasting the received signal power as a form aligning with a single-layer NN model. By this means, the CSI can be estimated by leveraging the backward propagation. Numerical results demonstrated that the proposed method can achieve a high CSI estimation accuracy for different group sizes. Moreover, a larger group size demands a larger number of TRPs and converges to a higher NMSE. The proposed method is expected to open up a new avenue for BD-RIS-aided wireless communications, enabling both high design flexibility and accurate channel estimation.

\end{document}